\documentclass[journal,10.5pt,final]{IEEEtran}

\usepackage{graphicx}
\usepackage{xcolor}
\usepackage{cite}
\usepackage{multirow}
\usepackage{rotate}
\usepackage{subfigure}
\usepackage{array}
\usepackage{url}

\DeclareGraphicsExtensions{.pdf,.png,.jpg}

\graphicspath{{figures/}}

\begin{document}

\title{Medical Cyber-Physical Systems Development: \\ A Forensics-Driven Approach}

\author{\IEEEauthorblockN{George Grispos$^{1}$, William Bradley Glisson$^{2}$, Kim-Kwang Raymond Choo$^{3}$} \\
	\IEEEauthorblockA{$^{1}$Lero -- The Irish Software Research Centre, University of Limerick, Limerick, Ireland \\
		$^{2}$School of Computing, University of South Alabama, Mobile, AL, USA \\ 
		$^{3}$Dept. of Information Systems and Cyber Security, The University of Texas at San Antonio, San Antonio, TX 78249, USA \\ 
		grisposg@acm.org, bglisson@southalabama.edu, raymond.choo@fulbrightmail.org}}


\maketitle

\begin{abstract} The synthesis of technology and the medical industry has partly contributed to the increasing interest in Medical Cyber-Physical Systems (MCPS). While these systems provide benefits to patients and professionals, they also introduce new attack vectors for malicious actors (e.g. financially- and/or criminally-motivated actors). A successful breach involving a MCPS can impact patient data and system availability. The complexity and operating requirements of a MCPS complicates digital investigations. Coupling this information with the potentially vast amounts of information that a MCPS produces and/or has access to is generating discussions on, not only, how to compromise these systems but, more importantly, how to investigate these systems. The paper proposes the integration of forensics principles and concepts into the design and development of a MCPS to strengthen an organization's investigative posture. The framework sets the foundation for future research in the refinement of specific solutions for MCPS investigations\footnote{This is the pre-print version of a paper presented at the 2nd International Workshop on Security, Privacy, and Trustworthiness in Medical Cyber-Physical Systems (MedSPT 2017). \\ \textbf{Recommended Citation}: \textit{Grispos G., Glisson W.B. and Choo K-K.R. 2017. Medical Cyber-Physical Systems Development: A Forensics-Driven Approach. In Proceedings of IEEE/ACM Conference on Connected Health: Applications, Systems and Engineering Technologies (CHASE 2017), pp. 108-114, Philadelphia, Pennsylvania, USA, 17-19 July. http://dx.doi.org/10.1109/CHASE.2017.48}}.

\end{abstract}


\IEEEpeerreviewmaketitle

\section{Introduction}

The fusion of information technology into the medical industry is creating opportunities and challenges for both practitioners and academicians. As technology continues to evolve, having up-to-date security and digital forensics capabilities is becoming increasingly important to the medical industry \cite{Wall16FDA}. For example, in a hospital context, what were previously being used as stand-alone medical devices are now being designed with embedded software and integrated with network interfaces \cite{king2013assuring,sokolsky2011challenges}. These network interfaces are used to communicate with other devices during patient treatments and healthcare systems that contain Electronic Health Records~\cite{king2013assuring,lee2010medical}. The integration of medical devices and healthcare systems is also referred to as a Medical Cyber-Physical System (MCPS) in the literature \cite{lee2010medical,lee2012challenges}. The implementation of a MCPS is expected to present several benefits that include increased safety, decreased false alarms and reduce workloads for medical professionals \cite{vasserman2012security}.

However, these benefits come with several developmental challenges \cite{sokolsky2011challenges,lee2010medical,lee2012challenges}. These challenges include a MCPS requiring a new design, verification and validation techniques, integrating interoperability, taking into account context-awareness, addressing security and privacy concerns and ensuring the MCPS can obtain safety certifications \cite{Chengetal2016,lee2010medical,lee2012challenges,MengLiXiangChoo2017,sokolsky2011challenges}. Moreover, unlike other safety-critical cyber-physical systems (e.g. aircraft), which are evaluated for safety by regulators before they are used \cite{Heetal2017}, a MCPS is often assembled at the patient's bedside \cite{king2013assuring}. As a result, the devices that contribute to a MCPS could realistically come from several vendors and that includes a variety of makes and models that were not developed to function together. These issues could result in additional attack vectors and future device recalls by regulators \cite{jiang2012cyber}.

To complicate matters further, a MCPS is likely to collect and manage large amounts of medical data \cite{kocabas2016emerging}. This will predictably increase security and privacy concerns within the healthcare community. A 2015 IBM report \cite{IBM16XForce} states that the healthcare industry is a high target industry. For example, reported attacks in 2015 resulted in the exposure of nearly 100 million medical records and the full extent of the threat landscape is unknown due to under- or non-reporting of attacks and cyber-security incidents \cite{IBM16XForce}. Echoing this idea, the Federal Bureau of Investigation (FBI) predicts that healthcare systems and medical devices are increasingly likely to become targets for attackers in the future \cite{FBI14Health}. The same FBI press release also emphasizes that the medical industry is not prepared to protect against basic cyber attacks \cite{FBI14Health}. Similar observations have also been made in more recent industry reports, indicating that most healthcare organizations are not prepared to handle security incidents and that only half of the surveyed organizations have a security incident response plan in place \cite{Ponemon15Criminal,Eset16State}. 

When a security incident occurs, organizations usually respond by conducting an investigation to establish the root cause of the incident and how it could be prevented in the future \cite{grisposrethinking}. In order to examine the causes of an incident, investigators rely on the residual data from systems, affected by the incident and supporting systems \cite{carrier2003getting}. However, such data might not always be available for a variety of reasons that include short data retention times, a lack of extraction capabilities and the costs associated with conducting such investigations~\cite{Stephenson200316,grispos2015security}. As a result, incident handlers may not be able to identify the causes of the security incident with any degree of confidence \cite{Stephenson200316}. Hence, there have been increasing calls for organizations to implement forensic-ready systems and infrastructure \cite{rowlingson2004ten,tan2001forensic}. Researchers have supported forensic readiness efforts by proposing that organizations implement policies and processes \cite{rowlingson2004ten}, align systems with forensics objectives \cite{reddy2013architecture} and the training of employees~\cite{tan2001forensic}.

In the past few years, researchers have proposed an alternative forensic readiness strategy called \textit{forensic-by-design} \cite{ab2016cloud,ab2016forensic,mink2016next}. Conceptually, forensic-by-design is similar to security-by-design, where requirements for forensics are integrated into relevant phases of the system development lifecycle, with the objective of developing forensic-ready systems \cite{ab2016forensic,Grispos17Ready}. In reality, a typical MCPS architecture could consist of different layers and contain multiple types of devices that could lead to several investigative issues. While there have only been a handful of known security incidents involving cyber-physical systems, a successful incident involving a MCPS could impact both system availability and patient data. The consequences of such incidents could, ultimately, result in the loss of life or physical disability. While a forensic-by-design approach will not stop a security incident from occurring, it can assist investigators in the examination of malicious or criminal activity involving a MCPS. For example, it can help with the preservation of evidential data, the analysis of an incident to determine root causes and accelerate the restoration of devices and services affected with an incident.

This paper proposes, through a conceptual framework, the integration of forensic-driven requirements into the MCPS design and development phases. The objective of this framework is to design and develop a MCPS that is driven with digital forensics in mind so that requirements for forensics are integrated into relevant phases of the development process. The paper is structured as follows. Section 2 examines challenges associated with the design and development of a MCPS as well as previous work in the forensic readiness domain. Sections 3 and 4 present the forensic-by-design framework and introduce forensic readiness testing, an approach for verifying and validating forensic-by-design approaches. Section 5 evaluates the framework through a hypothetical case study and the last section presents the conclusions and ideas for future work.

\section{Related Work}

Researchers are continuously demonstrating that devices are at risk in a medical context \cite{luckett2017attack,van2016identifying,glisson2015compromising}. Coupling these activities with research indicating that there is increasing interest in residual data in a general legal context \cite{berman2015investigating,mcmillan2013investigating} stimulates governmental, practitioner and academic interest in appropriate regulatory issues and medical investigation capabilities. However, a growing number of researchers are arguing that, due to their size and complexity, relative to traditional medical systems, medical cyber-physical systems present several development challenges \cite{king2013assuring,lee2012challenges, jiang2012cyber,kocabas2016emerging,venkatasubramanian2012security}. The term \textit{Medical Cyber-Physical System} (MCPS) is defined as a ``safety-critical, interconnected, intelligent system of medical devices'' \cite{lee2012challenges}.

Lee and Sokolsky \cite{lee2010medical} suggest that a MCPS integrates the digital, physical and medical worlds and go on to argue that the development of such systems introduces significant challenges for the information technology, medical and regulatory communities. These challenges include deploying a MCPS so that they can be integrated into custom clinical scenarios, using model-based development approaches to assess patient safety before deployment, incorporating verification and certification early on in the development process, and finding the balance between security and flexibility \cite{lee2010medical,lee2012challenges}. King, et al. \cite{king2013assuring} add that unlike most other safety-critical cyber-physical systems, a MCPS is likely to be constructed at the patient's bedside. As a result, this can introduce additional safety challenges if regulators will not know the brand, make or model of the devices used in the particular MCPS instance~\cite{king2013assuring}. Other researchers have raised similar concerns. Sokolsky, et al. \cite{sokolsky2011challenges} add that regulators could face several challenges when attempting to approve modern medical devices that could form part of a MCPS. The authors go on to state that communication within a MCPS will not only introduce network failure concerns, but will also introduce additional security and privacy concerns. Venkatasubramanian, et al. \cite{venkatasubramanian2012security} examined the security challenges and research direction in MCPSs, and indicated that the interoperability of devices in a MCPS is creating a much larger attack surface for malicious actors. They also pointed out that MCPSs provide a unique set of security challenges that are distinct from other types of cyber-physical systems. 

Growing security concerns with medical devices has prompted increased regulatory pressure to be applied to organizations in the healthcare industry mandating security incident handling capabilities, including forensic investigations \cite{johnson2003handbook}. Forensic investigations are often part of an organization's security incident response capability, with the aim of attempting to establish the six key questions of an incident: what, why, who, when, where and how \cite{freiling2007common}. While the objective of security incident response is to restore service and learn about the causes of a security incident, digital forensics is concerned with the collection and analysis of evidential data, which can then be used as evidence in court \cite{freiling2007common}. Typically, security incident handlers within an organization collect and analyze potential evidential data after a security incident has occurred. As a result, there are concerns that organizations appear to be complacent with data activities prior to an incident \cite{rowlingson2004ten}. Hence, data that is required for an investigation will either exist and is preserved by a system or it does not exist and this can hinder an effective investigation \cite{rowlingson2004ten}. These concerns prompted suggestions that organizations need to be more proactive, in reference to digital forensics, and structure environments to retain data required for investigations \cite{tan2001forensic}. This approach or stance is known as forensic readiness  \cite{tan2001forensic}. Previous forensic readiness research focused on the implementation of policies and processes, aligning systems with forensic objectives and the impact and training of employees \cite{rowlingson2004ten, barske2010digital}.

Within the information security domain, researchers have proposed forensics-based policies that can provide an organized structure within organizations \cite{grisposcloud,barske2010digital,endicott2007theoretical}. Barske, et al. \cite{barske2010digital} state that these policies should define how an organization will monitor their systems, the conditions where data will be preserved for an investigation and the development of policies that are needed to define when an investigation must be undertaken. Supporting these suggestions, Rowlingson \cite{rowlingson2004ten} adds that policies should also define the identification of evidential data sources and prompt the secure storage of any data that could be required for an investigation. Like security policies, Endicott-Popovsky, et al. \cite{endicott2007theoretical} argue that policies concerning forensic readiness should be regularly audited to ensure continuous readiness. 

While some researchers have proposed using policies, others have argued that well-defined processes can enhance organizational forensic readiness \cite{tan2001forensic,reddy2009forensic,yasinsac2001policies}. For example, supporting investigators with a well-defined investigation process can help reduce any mistakes during an investigation \cite{yasinsac2001policies}. Similarly, the availability of data preservation and collection processes can increase the speed of an investigation and decrease the cost of reacquiring any compromised evidential data \cite{rowlingson2004ten}. Separately, other researchers have proposed that organizations implement processes to guide log acquisition along with specific extraction for investigations \cite{tan2001forensic,endicott2007theoretical}. However, there is minimal research addressing comprehensive design considerations in terms of a MCPS. 

\section{Forensic-By-Design Framework for MCPS}
\label{sec:FBD}

\begin{figure*}
\centering
\includegraphics[trim=-180 170 40 40, scale=0.45]{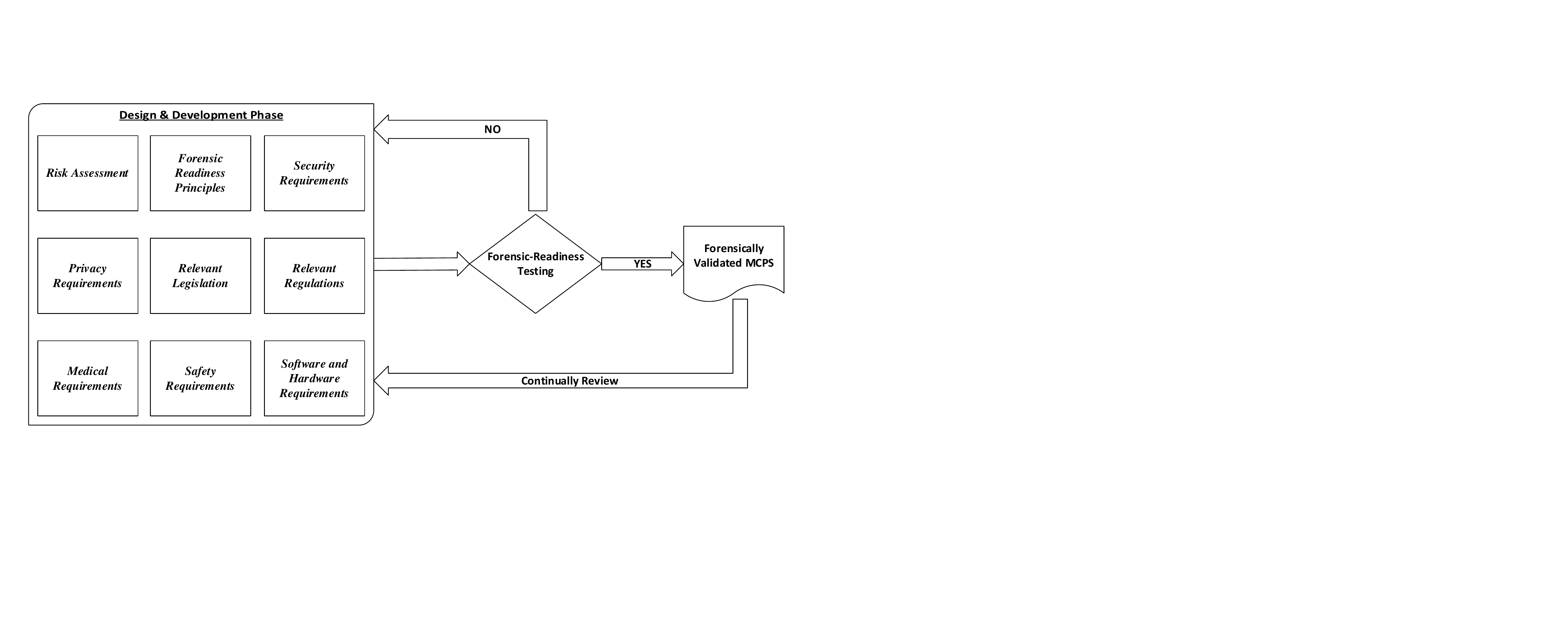}
\caption{MCPS forensic-by-design framework}
\label{MCPSFramework}
\end{figure*}

The forensic-by-design framework for a MCPS consists of nine components as shown in Figure \ref{MCPSFramework}. The components are presented as a black-box with the objective of designing a forensic-ready MCPS that can assist developers in the overall design and development process along with guiding investigators in the examination of security incidents. 

\vspace{-3pt}
\subsection{Risk Assessment}

Organizations do not have infinite resources (e.g. financial and manpower) and risk assessments are a key component of an organization's effort to identify threats and vulnerabilities that can affect systems and applications. Integrating risk assessments into the design and development of a MCPS provides two purposes. It can be used to identify risks that are possible within a MCPS and prompt the implementation of security controls to reduce this risk. However, residual risk could persist and this, in turn, could result in security incidents based on the threats and vulnerabilities. Therefore, MCPS stakeholders can also use a risk assessment to identify potential incidents that can occur within a MCPS environment and then apply forensic readiness principles to these incidents to support investigations, should an incident occur. 

\vspace{-3pt}
\subsection{Forensic Readiness Principles}
\label{sub:readiness}

Forensic readiness focuses on an organization's ability to maximize the collection of evidential data, while minimizing the cost of an investigation \cite{rowlingson2004ten,tan2001forensic}. This framework adopts the following forensic readiness activities originally proposed by Rowlingson \cite{rowlingson2004ten}:

\begin{itemize}
	\item \textbf{Define the business scenarios that require digital evidence}. MCPS stakeholders need to identify the various scenarios where digital evidence could be required based on the results of the risk assessment. In a healthcare context, MCPS stakeholders may also need to provide evidence to demonstrate regulatory compliance and/or use this information to prove the impact of an incident. 
	
	\item \textbf{Identify available sources and different types of potential evidence}. Kocabas, et al. \cite{kocabas2016emerging} argue that a MCPS architecture consists of four layers (data acquisition, data concentration, cloud processing and action layers). Therefore, evidence sources within a MCPS could include wearable sensors, medical applications and devices in the data acquisition layer; cloudlets and gateway servers in the data concentration layer; public and private cloud environments in the cloud processing layer and the outcomes of activities taken in the action layer.
	
	\item \textbf{Determine the evidence collection requirement}. By default, the various devices, applications, services and systems that constitute a MCPS are, typically, not designed with forensics in mind. Hence, strategies need to be identified and implemented on how potential evidence will be collected. For example, body-sensors are unlikely to have the storage capacity to store event metadata from the specific sensor. As a result, cloudlets or gateway devices connected to the sensor need to collect and store event metadata in a database which can then examined, should the need arise.
	
	\item \textbf{Establish a policy for secure storage and handling of potential evidence}. If evidential data collected from a MCPS is to be used during an investigation, it needs to be handled and stored correctly \cite{casey2011digital}. One solution could involve the use of a centralized secure database that can be used to store evidential data. Furthermore, investigators tasked to investigate incidents in a MCPS must follow evidence handling procedures that adhere to digital forensics standards.
	
	\item \textbf{Specify circumstances when escalation to a full formal investigation should be launched}. The complexity of a MCPS could inevitably make it difficult to identify when a suspicious event has occurred. However, when a suspicious event has been identified, an organization will need to make use of an incident response taxonomy so that it can `grade' the severity or impact of the event~\cite{grispos2016enhancement}. The outcome of this `grading' taxonomy provides an indication as to organization's need to escalate to a full incident response investigation.  
	
	\item \textbf{Train staff in incident awareness, so that all those involved understand their role in the digital evidence process and the legal sensitivities of evidence}. Grispos, et al. \cite{Grispos17SIRR} and Christopher, et al. \cite{christopher2016honeypots} identified that there are opportunities to improve security incident awareness and reporting practices within organizations. With a magnitude of potential stakeholders involved in a MCPS, similar opportunities could exist in MCPS contexts. Furthermore, organizations need to define practices and policies that can be used in a MCPS that will promote an environment that is conducive to protecting evidential data, particularly, when this data could be used in court.
\end{itemize}

\vspace{-4pt}
\subsection{Security Requirements}

When stakeholders establish that a system provides some value to an organization, they will attempt to protect this system from harm by expressing security requirements during its development \cite{haley2008security}. From a forensic-by-design perspective, a failure to correctly implement these requirements could result in the need to conduct a forensic investigation. However, investigation findings and knowledge gained from responding to security incidents could also be used to delineate better security requirements in the context of a MCPS. The various applications, devices and systems that constitute a MCPS will, in all likelihood, require different security requirements. Therefore, obtaining a better understanding of the security requirements needed for this complex interconnected system will improve protection initiatives for a MCPS.

\vspace{-4pt}
\subsection{Privacy Requirements}

Sensors and devices within a MCPS, potentially, collect highly sensitive information. For example, some sensors within a MCPS could act as `lab-on-a-chip' and can detect the presence of medical drugs, record temperatures and collect bio-markers \cite{burleson2012design}. These sensors and devices require stakeholders to take into consideration a special set of privacy-preserving requirements. For instance, private patient information generated from general use of a MCPS could be exposed to forensic investigators examining a security incident involving a malicious actor. Privacy concerns indicate that designing a forensic-driven MCPS will likely require acknowledgement that privacy requirements may conflict with forensic strategies. 

\vspace{-3pt}
\subsection{Relevant Legislation}

Mission-critical software, such as a MCPS, often needs to be designed to comply with a variety of laws \cite{garcia2016towards}. Therefore, the design of a forensic strategy for a MCPS needs to take into consideration relevant healthcare legislation that impacts the system. The system also needs to ensure that it takes into consideration privacy legislation for individuals interfacing with the system. These will differ from country to country. An example of this issue is visible in through data protection legislation in the UK and the US. UK legislation is very stringent when compared to the, overall, US legislation on the topic. Existing legislation in different countries will also impact evidential collection, storage and processing which could perceivably impact MCPS stakeholders. With the integration of cloud storage capabilities into healthcare environments, it is conceivable that healthcare data could be acquired in one country and legitimately processed in another country. When a forensic investigation needs to take place in this scenario, conflicts could arise in terms of data protection, data acquisition and retention. Hence, relevant legislation in these areas will need to be taken into consideration. 

\vspace{-3pt}
\subsection{Relevant Regulations}

In addition to considering relevant laws, healthcare regulations and standards should also be considered. Healthcare regulations influence the monitoring, collection and storage of potential evidential data that is collected from a MCPS. For example, Section 164.308 of the `Security Rule' part of the Health Insurance Portability and Accountability Act (HIPAA) of 1996 \cite{USGov96HIPAA} specifically requires organizations to ``implement policies and procedures to prevent, detect, contain, and correct security violations'' \cite{USGov07Sec}. The Security Rule also indicates that digital evidence is preserved and that organizations document incidents that occur as well as their outcomes \cite{USGov07Sec}. 

\vspace{-4pt}
\subsection{Medical Requirements}

Understanding the interaction between a medical device and its various medical environments is crucial within a MCPS. For the purpose of this discussion, medical requirements describe the requirements that need to be satisfied in order for an application, device or system to have a medical purpose. These requirements drive the development of a device's functionality like being able to push insulin to a patient from an infusion pump. It is worth noting that diverse medical requirements for MCPS components could conflict with a forensic-driven MCPS development approach. For example, in the event of an incident or breach, medical requirements will likely demand that a component delivers its medical functionality. However, this could conflict with the preservation and collection of evidential data. Hence, stakeholders need to first identify the medical requirements for a MCPS and then perform a trade-off analysis between medical requirements with forensics-related requirements that support forensic readiness.

\vspace{-4pt}
\subsection{Safety Requirements}

Interconnected medical devices that can form part of a MCPS need to be designed and validated with patient safety in mind \cite{lee2010medical}. The inclusion of forensic strategies into the development of a MCPS could introduce additional safety challenges. This is because many of the existing forensic strategies and incident response process focus on investigations involving office-based systems and provide minimal support for investigations involving safety-critical system such as a MCPS \cite{johnson2003handbook}. For example, traditional approaches advocate the shutting down of systems and preservation of evidence contained in device memory. However, shutting down a safety-critical application may be impossible (e.g. a fusion pump or pacemaker) to collect potential evidential metadata. Therefore, MCPS stakeholders need to take into consideration safety requirements when designing and developing the forensic strategy that will be integrated into the design of their MCPS, as well as considering other alternatives such as live forensic acquisition techniques.  

\vspace{-3pt}
\subsection{Software and Hardware Requirements}

A MCPS will likely contain a variety of different applications, devices and systems, all running different hardware configurations and software developed by different vendors. The reliance between these various components within a MCPS inevitably complicates the preservation and collection of evidential data. As noted in Section \ref{sub:readiness}, potential evidential data could exist in various layers of a MCPS architecture. Therefore, understanding the various hardware and software requirements that will emerge from each layer supports the identification, preservation and collection of data that could be used in a forensic investigation. For example, if a mobile device is used as a gateway between a monitoring sensor and a cloud storage facility, a forensic investigator needs to identify potential evidence that can be extracted from the mobile device. This perceivably includes addressing the challenges of collecting evidence from a cloud computing environment. 

\vspace{-4pt}
\section{Forensic-Readiness Testing}

After a prototype MCPS has been designed and developed using a forensic-driven approach, it is critical to test and confirm the forensic principles that have been integrated into the relevant phases. Effectively, testing and validating forensic capabilities provides MCPS stakeholders with information about the ability to conduct investigations in a MCPS environment. Furthermore, several researchers have argued that an effective software verification process is critical for the certification of medical device software \cite{king2013assuring, sokolsky2011challenges,lee2010medical,lee2012challenges}. However, there is no current standard for testing, validating and verifying forensic-by-design software and systems. Hence, the following activities are proposed as a starting point for addressing this deficiency within a MCPS forensic-driven development approach. These include: 

\begin{itemize}
	\item Verifying that evidential data is created within a MCPS and that this evidential data satisfies the various business scenarios that could require digital evidence. Furthermore, any evidential data that is created and stored within a MCPS satisfies legal, regulatory and industry demands and other requirements (e.g. safety requirements).
	
	\item Ensuring that potential evidence can be preserved and collected both reliably and compliant with digital forensics standards. This includes evaluating security incident and evidence handling procedures to ensure that they adhere to digital forensics standards.
	
	\item Ensuring that various components of a MCPS are `forensically sound' and do not contaminate, modify or delete any evidential data that could be used in a forensic investigation of a MCPS. 
	
	\item Performing a safety, security and privacy requirements check against the MCPS to ensure that they have been satisfied. Moreover, any risks and trade-offs between these requirements and those related to forensics have been sufficiently acknowledged and/or mitigated. 

\end{itemize}

In the US, to submit digital evidence into a court it must be both reliable and relevant \cite{grispos2015security}. The reliability of digital evidence is tested by applying the Daubert standard \cite{USSC}. The Daubert standard proposes four queries to assess the reliability of digital evidence:

\begin{itemize}
	\item Can and has the procedure used to acquire the data been tested?
	\item Is there a known error rate for this procedure? 
	\item Has the procedure been published and subject to peer-review?
	\item Does the relevant community accept the procedure as an approach for acquiring the data? 

\end{itemize}

These four queries from the Daubert standard could serve as an additional approach to validating and verifying forensic-by-design software. A viable approach to utilizing these queries is to develop assurance or test cases that will help answer the above questions. In this scenario, software testers, security incident response personnel and forensic investigators could write test cases to determine error rates for evidence acquisition procedures. This would likely involve preloading a known data set into a MCPS and then using a forensic tool to acquire the data from the MCPS. After analyzing the acquired data, the number of recovered files can be compared to the known data set and the differences used to establish error rates of the procedure and the tool with the device in various states.

\begin{table*}[!h]
	\begin{center}
		\begin{tabular}{| c | p{13cm} | }
			\hline
			\multicolumn{1}{|c|}{\textbf{Design and Development}} &  \multicolumn{1}{|c|}{\textbf{Potential Actions}} \\ \hline
			
			Risk Assessment & Undertake a risk assessment to identify, assess, evaluate and prioritize threats and vulnerabilities that could impact the MCPS. For example, a risk involving a man-in-the-middle attack could be identified in the network connection between the BCMA and the fusion pump. Such a risk would be classified as high-priority because such an attack could increase or decrease the functionality presented by the fusion pump. \\ \hline
			
			Forensic Readiness & Information identified from the risk assessment can be used to drive the forensic readiness practices in this phase. For example, after identifying that a man-in-the-middle attack is possible (a business scenario that could require evidential data), stakeholders need to identify what sources of evidence can exist as well as what type of evidence is needed to investigate the incident. This involves examining all the layers within the MCPS architecture, along with the network communication layer. 
			
			In addition, evidence and incident handling best practices concerning this type of evidence (e.g. network-based evidence) must be identified and incident handlers should be trained on how to preserve, collect and store this evidence according to relevant standards. \\ \hline
			
			Security Requirements & Using the risk assessment, stakeholders can express what security requirements are needed to protect the MCPS from harm. In addition, stakeholders can consider the domain knowledge to identify information from investigations of previous similar threats and propose better security requirements for the context of the MCPS. \\ \hline
			
			Privacy Requirements & The interactions between the fusion pump and the BCMA generate medical data that is then transferred to relevant hospital systems. As a result, privacy requirements need to be identified to ensure that a forensic investigation (e.g. to investigate the man-in-the-middle incident) does not expose private patient information when attempting to identify the malicious actor or actors involved in the incident. This could mean that data is anonymized to the point that the forensic investigator does not recognize the patient using the fusion pump.  \\ \hline
			
			Relevant Legislation & Privacy legislation needs to be followed when designing and examining interactions in the MCPS. In the UK, relevant privacy legislation that would need to be considered from a compliance perceptive is the Data Protection Act and, in the US, data privacy is addressed, to a large extent, in the Health Insurance Portability and Accountability Act. \\ \hline
			
			Relevant Regulations & Investigations need to ensure that they are compliant with regulations. In the current scenario, if an investigation revealed that there were problems with the pump or a connected device, then these issues would need to be reported, in compliance with mandatory or voluntary regulations, to the Food and Drug Administration (FDA) for United States based institutions. \\ \hline
			
			Medical Requirements & The requirements that focus on pump operations like a pump continuing or discontinuing an infusion, for a specific patient, based on the information gathered from the interaction between the BCMA and the hospital systems is a vital medical requirement. However, these requirements also need to be considered from a forensics perspective. For example, the date and time the infusion pump was started/stopped, as well as where the command to initiate this action came from could be useful information to a forensic investigator. \\ \hline
			
			Safety Requirements & A safety analysis needs to be undertaken to identify, depending on the type of infusion pump and the medication being dispersed, whether it is safe to conduct a forensic investigation whilst the pump is still providing a medical function to a patient. The results of this analysis would then determine if the pump is powered on or off before a forensic investigation can take place. Furthermore, stakeholders may need to consider the impact on evidential metadata should the pump be turned off, as this could erase temporary information of value.  \\ \hline
			
			Software \& Hardware Requirements & Interoperability between the various MCPS hardware and software components is evident in the case study example. From a forensics perspective, stakeholders need to understand what protocols are in use between the BCMA and the hospital system as well as how the software components have been designed and their impact on evidence preservation and collection. \\ \hline
		\end{tabular}
	\end{center}
	\caption{Conceptual Framework Actions}
	\label{tab:Actions}
\end{table*} 

Any problems that emerge during testing, validation and verification will involve refining some of the requirements presented in Section \ref{sec:FBD} and then re-testing associated factors. After testing, a MCPS should be forensic-ready based on the requirements defined in the previous sections.

\section{Hypothetical Case Study}

A plausible case study example would be an integrated fusion pump. In this scenario, a scanner would be used to scan a patient's identification. This information would then be transferred to the fusion pump. The pump conceivably has a network connection to utilize a Bar Coded Medication Administration (BCMA) control system \cite{Rouse15BCMA}. The idea behind a BCMA is to validate medication distribution with existing orders for specific patients \cite{Rouse15BCMA}. Once, this has been validated, the infusion is initiated by a medical professional or by the pump itself. The medical data generated from this interaction is then transferred to relevant hospital systems for analysis by the next shift of medical professionals, pharmaceutical inventory systems, billing systems, general reporting systems and archival systems. A sample implementation of the conceptual forensic-by-design framework is presented in Table \ref{tab:Actions}.

Each of the systems that are involved in this example would need to be examined from the perspective of the nine design and development phases. Once these phases have been completed, the system in its entirety would need to pass forensic-readiness testing. Once this has been completed, a document would be produced certifying that specific MCPS configuration from a forensic readiness perspective. If the configuration changes, the process would need to be repeated for re-certification. 

\section{Conclusions and Future Work}

As cyber-physical systems continue to integrate into medical environments, the need for forensic investigations will continue to escalate. The forensic-by-design framework presented in this paper provides a starting point for conversations, research and solutions that could be used to address this issue. 

Future work will endeavor to develop practical solutions that will address specific technical and procedural implementations of the nine design and development stages identified in the conceptual framework in a MCPS context. In addition, the solution(s) should ideally be implemented and evaluated using a test-bed at a hospital setting so that the solution(s) can be refined, if necessary.  

\section*{Acknowledgments}
The first author was partially supported by SFI Grant No. 13/RC/2094 and ERC Advanced Grant. No. 291652 (ASAP), and the last author is supported by the Cloud Technology Endowed Professorship. 

\bibliographystyle{IEEEtran}
\bibliography{MCPS}

\begin{thebibliography}{10}
\providecommand{\url}[1]{#1}
\csname url@samestyle\endcsname
\providecommand{\newblock}{\relax}
\providecommand{\bibinfo}[2]{#2}
\providecommand{\BIBentrySTDinterwordspacing}{\spaceskip=0pt\relax}
\providecommand{\BIBentryALTinterwordstretchfactor}{4}
\providecommand{\BIBentryALTinterwordspacing}{\spaceskip=\fontdimen2\font plus
\BIBentryALTinterwordstretchfactor\fontdimen3\font minus
  \fontdimen4\font\relax}
\providecommand{\BIBforeignlanguage}[2]{{%
\expandafter\ifx\csname l@#1\endcsname\relax
\typeout{** WARNING: IEEEtran.bst: No hyphenation pattern has been}%
\typeout{** loaded for the language `#1'. Using the pattern for}%
\typeout{** the default language instead.}%
\else
\language=\csname l@#1\endcsname
\fi
#2}}
\providecommand{\BIBdecl}{\relax}
\BIBdecl

\bibitem{Wall16FDA}
{The Wall Street Journal}, ``Fda approves world's smallest pacemaker that
  attaches directly to heart.'' Available:
  http://www.foxnews.com/health/2016/04/07/fda-approves-worlds-smallest-pacemaker-that-attaches-directly-to-heart.html,
  2016.

\bibitem{king2013assuring}
A.~L. King, L.~Feng, O.~Sokolsky, and I.~Lee, ``Assuring the safety of
  on-demand medical cyber-physical systems,'' in \emph{1st International
  Conference on Cyber-Physical Systems, Networks and Applications}.\hskip 1em
  plus 0.5em minus 0.4em\relax IEEE, 2013, pp. 1--6.

\bibitem{sokolsky2011challenges}
O.~Sokolsky, I.~Lee, and M.~Heimdahl, ``Challenges in the regulatory approval
  of medical cyber-physical systems,'' in \emph{9th ACM international
  conference on Embedded software}.\hskip 1em plus 0.5em minus 0.4em\relax ACM,
  2011, pp. 227--232.

\bibitem{lee2010medical}
I.~Lee and O.~Sokolsky, ``Medical cyber physical systems,'' in \emph{47th
  ACM/IEEE Design Automation Conference}.\hskip 1em plus 0.5em minus
  0.4em\relax IEEE, 2010, pp. 743--748.

\bibitem{lee2012challenges}
I.~Lee, O.~Sokolsky, S.~Chen, J.~Hatcliff, E.~Jee, B.~Kim, A.~King,
  M.~Mullen-Fortino, S.~Park, A.~Roederer \emph{et~al.}, ``Challenges and
  research directions in medical cyber--physical systems,'' \emph{Proceedings
  of the IEEE}, vol. 100, no.~1, pp. 75--90, 2012.

\bibitem{vasserman2012security}
E.~Y. Vasserman, K.~K. Venkatasubramanian, O.~Sokolsky, and I.~Lee, ``Security
  and interoperable-medical-device systems, part 2: Failures, consequences, and
  classification,'' \emph{IEEE security \& privacy}, vol.~10, no.~6, pp.
  70--73, 2012.

\bibitem{Chengetal2016}
C.~Guo, R.~Zhuang, Y.~Jie, Y.~Ren, T.~Wu, and K.-K.~R. Choo, ``Fine-grained
  database field search using attribute-based encryption for e-healthcare
  clouds,'' \emph{Journal of Medical Systems}, vol.~40, p. 235, 2016.

\bibitem{MengLiXiangChoo2017}
W.~Meng, W.~Li, Y.~Xiang, and K.-K.~R. Choo, ``A bayesian inference-based
  detection mechanism to defend medical smartphone networks against insider
  attacks,'' \emph{Journal of Network and Computer Applications}, vol.~78, pp.
  162---169, 2017.

\bibitem{Heetal2017}
D.~He, N.~Kumar, K.-K.~R. Choo, and W.~Wu, ``Efficient hierarchical
  identity-based signature with batch verification for automatic dependent
  surveillance-broadcast system,'' \emph{IEEE Transactions on Information
  Forensics and Security}, vol.~12, pp. 454---464, 2017.

\bibitem{jiang2012cyber}
Z.~Jiang, M.~Pajic, and R.~Mangharam, ``Cyber--physical modeling of implantable
  cardiac medical devices,'' \emph{Proceedings of the IEEE}, vol. 100, no.~1,
  pp. 122--137, 2012.

\bibitem{kocabas2016emerging}
O.~Kocabas, T.~Soyata, and M.~K. Aktas, ``Emerging security mechanisms for
  medical cyber physical systems,'' \emph{IEEE/ACM Trans. on Comp. Biology and
  Bioinformatics}, vol.~13, no.~3, pp. 401--416, 2016.

\bibitem{IBM16XForce}
IBM, ``Ibm x-force threat intelligence index,'' IBM, Tech. Rep., 2016.

\bibitem{FBI14Health}
{FBI Cyber Division}, ``(u) health care systems and medical devices at risk for
  increased cyber intrusions for financial gain,'' 2014.

\bibitem{Ponemon15Criminal}
{Ponemon Institute}, ``Criminal attacks are now leading cause of data breach in
  healthcare, according to new ponemon study.'' Available:
  \url{http://www.ponemon.org/news-2/66}, 2015.

\bibitem{Eset16State}
Eset, ``The state of cybersecurity in healthcare organizations,'' 2016.

\bibitem{grisposrethinking}
G.~Grispos, W.~B. Glisson, and T.~Storer, ``Rethinking security incident
  response: The integration of agile principles,'' in \emph{20th Americas
  Conference on Information Systems, Savannah, Georgia, USA}, 2014.

\bibitem{carrier2003getting}
B.~Carrier and E.~H. Spafford, ``Getting physical with the digital
  investigation process,'' \emph{Int. Journal of Digital Evidence}, vol.~2,
  no.~2, pp. 1--20, 2003.

\bibitem{Stephenson200316}
P.~Stephenson, ``Conducting incident post mortems,'' \emph{Computer Fraud and
  Security}, vol. 2003, no.~4, pp. 16 -- 19, 2003.

\bibitem{grispos2015security}
G.~Grispos, W.~Glisson, and T.~Storer, ``Security incident response criteria: A
  practitioner's perspective,'' in \emph{21st Americas Conference on
  Information Systems, Puerto Rico, USA}, 2015.

\bibitem{rowlingson2004ten}
R.~Rowlingson, ``A ten step process for forensic readiness,''
  \emph{International Journal of Digital Evidence}, vol.~2, no.~3, pp. 1--28,
  2004.

\bibitem{tan2001forensic}
J.~Tan, ``Forensic readiness,'' Cambridge, MA:@ Stake, Tech. Rep., 2001.

\bibitem{reddy2013architecture}
K.~Reddy and H.~S. Venter, ``The architecture of a digital forensic readiness
  management system,'' \emph{Comp. \& Sec.}, vol.~32, pp. 73--89, 2013.

\bibitem{ab2016cloud}
N.~H. Ab~Rahman, N.~D.~W. Cahyani, and K.-K.~R. Choo, ``Cloud incident handling
  and forensic-by-design: cloud storage as a case study,'' \emph{Concurrency
  and Computation: Practice and Experience}, 2016.

\bibitem{ab2016forensic}
N.~H. Ab~Rahman, W.~B. Glisson, Y.~Yang, and K.-K.~R. Choo,
  ``Forensic-by-design framework for cyber-physical cloud systems,'' \emph{IEEE
  Cloud Computing}, vol.~3, no.~1, pp. 50--59, 2016.

\bibitem{mink2016next}
D.~Mink, A.~Yasinsac, K.-K.~R. Choo, and W.~Glisson, ``Next generation aircraft
  architecture and digital forensic,'' in \emph{22nd Americas Conference on
  Information Systems, San Diego, USA}, 2016.

\bibitem{Grispos17Ready}
G.~Grispos, J.~Garc{\'\i}a-Gal{\'a}n, L.~Pasquale, and B.~Nuseibeh, ``Are you
  ready? towards the engineering of forensic-ready systems,'' in \emph{11th
  IEEE International Conference on Research Challenges in Information Science},
  2017, \textit{In Press}.

\bibitem{luckett2017attack}
P.~Luckett, J.~McDonald, and W.~Glisson, ``Attack-graph threat modeling
  assessment of ambulatory medical devices,'' in \emph{Proceedings of the 50th
  Hawaii International Conference on System Sciences}, 2017.

\bibitem{van2016identifying}
M.~Van~Devender, W.~Glisson, M.~Campbell, and M.~Finan, ``Identifying
  opportunities to compromise medical devices,'' in \emph{22nd Americas
  Conference on Information Systems, San Diego, USA}, 2016.

\bibitem{glisson2015compromising}
W.~B. Glisson, T.~Andel, T.~McDonald, M.~Jacobs, M.~Campbell, and J.~Mayr,
  ``Compromising a medical mannequin,'' in \emph{21st Americas Conference on
  Information Systems, Puerto Rico, USA}, 2015.

\bibitem{berman2015investigating}
K.~J. Berman, W.~B. Glisson, and L.~M. Glisson, ``Investigating the impact of
  global positioning system evidence,'' in \emph{48th Hawaii International
  Conference on System Sciences}, 2015.

\bibitem{mcmillan2013investigating}
J.~E.~R. McMillan, W.~B. Glisson, and M.~Bromby, ``Investigating the increase
  in mobile phone evidence in criminal activities,'' in \emph{46th Hawaii
  International Conference on System Sciences}, 2013.

\bibitem{venkatasubramanian2012security}
K.~K. Venkatasubramanian, E.~Y. Vasserman, O.~Sokolsky, and I.~Lee, ``Security
  and interoperable-medical-device systems, part 1,'' \emph{IEEE security \&
  privacy}, vol.~10, no.~5, pp. 61--63, 2012.

\bibitem{johnson2003handbook}
C.~Johnson, \emph{A Handbook of Incident \& Accident Reporting}.\hskip 1em plus
  0.5em minus 0.4em\relax Glasgow Univ. Press, 2003.

\bibitem{freiling2007common}
F.~Freiling and B.~Schwittay, ``A common process model for incident response
  and digital forensics,'' in \emph{International Conference on IT Security
  Incident Management and IT Forensics 2007}, 2007.

\bibitem{barske2010digital}
D.~Barske, A.~Stander, and J.~Jordaan, ``A digital forensic readiness framework
  for south african sme's,'' in \emph{Information Security for South Africa
  (ISSA), 2010}.\hskip 1em plus 0.5em minus 0.4em\relax IEEE, 2010, pp. 1--6.

\bibitem{grisposcloud}
G.~Grispos, W.~B. Glisson, and T.~Storer, ``{Cloud Security Challenges:
  Investigating Policies, Standards, And Guidelines In A Fortune 500
  Organization},'' in \emph{21st European Conference on Info. Systems}, 2013.

\bibitem{endicott2007theoretical}
B.~Endicott-Popovsky, D.~A. Frincke, and C.~A. Taylor, ``A theoretical
  framework for organizational network forensic readiness.'' \emph{JCP},
  vol.~2, no.~3, pp. 1--11, 2007.

\bibitem{reddy2009forensic}
K.~Reddy and H.~Venter, ``A forensic framework for handling information privacy
  incidents,'' in \emph{Int. Conf. on Digital Forensics}.\hskip 1em plus 0.5em
  minus 0.4em\relax Springer, 2009.

\bibitem{yasinsac2001policies}
A.~Yasinsac and Y.~Manzano, ``Policies to enhance computer and network
  forensics,'' in \emph{Workshop on Info. Assurance and Sec.}, 2001, pp.
  289--295.

\bibitem{casey2011digital}
E.~Casey, \emph{Digital evidence and computer crime: Forensic science,
  computers, and the internet}.\hskip 1em plus 0.5em minus 0.4em\relax Academic
  press, 2011.

\bibitem{grispos2016enhancement}
G.~Grispos, ``On the enhancement of data quality in security incident response
  investigations,'' Ph.D. dissertation, Univ. of Glasgow, 2016.

\bibitem{Grispos17SIRR}
G.~Grispos, W.~B. Glisson, D.~Bourie, and T.~Storer, ``{Security Incident
  Recognition and Reporting (SIRR): An Industrial Perspective},'' in \emph{23rd
  Americas Conference on Info. Systems, Boston, USA}, 2017, \textit{In Press}.

\bibitem{christopher2016honeypots}
L.~Christopher, K.-K.~R. Choo, and A.~Dehghantanha, ``Honeypots for employee
  information security awareness and education training: a conceptual easy
  training model,'' in \emph{Contemporary Digital Forensic Investigations of
  Cloud and Mobile Applications}, K.-K.~R. Choo and D.~A, Eds.\hskip 1em plus
  0.5em minus 0.4em\relax Syngress, an Imprint of Elsevier, 2016, pp. 111--129.

\bibitem{haley2008security}
C.~Haley, R.~Laney, J.~Moffett, and B.~Nuseibeh, ``Security requirements
  engineering: A framework for representation and analysis,'' \emph{IEEE
  Transactions on Software Engineering}, vol.~34, no.~1, pp. 133--153, 2008.

\bibitem{burleson2012design}
W.~Burleson, S.~S. Clark, B.~Ransford, and K.~Fu, ``Design challenges for
  secure implantable medical devices,'' in \emph{Proceedings of the 49th Annual
  Design Automation Conference}.\hskip 1em plus 0.5em minus 0.4em\relax ACM,
  2012, pp. 12--17.

\bibitem{garcia2016towards}
J.~Garc{\'\i}a-Gal{\'a}n, L.~Pasquale, G.~Grispos, and B.~Nuseibeh, ``Towards
  adaptive compliance,'' in \emph{11th Int. Symposium on Software Engineering
  for Adaptive and Self-Managing Systems}.\hskip 1em plus 0.5em minus
  0.4em\relax ACM, 2016, pp. 108--114.

\bibitem{USGov96HIPAA}
{U.S. Goverment}, ``The health insurance portability and accountability act,
  pub.l. 104-191,'' 1996.

\bibitem{USGov07Sec}
{U.S. Dept. of Health \& Human Services}, ``Security standards: Administrative
  safeguards,'' 2007.

\bibitem{USSC}
{U.S. Supreme Court}, ``{Daubert v. Merrell Dow Pharmaceuticals},'' 1993.

\bibitem{Rouse15BCMA}
M.~Rouse, ``Bar coded medication administration (bcma),'' Available:
  http://searchhealthit.techtarget.com/definition/Bar-Coded-Medication-Administration,
  2015.

\end{thebibliography}

\end{document}